\pdfoutput=1
\documentclass[10pt]{article}

\usepackage{algorithm}
\usepackage{color}
\usepackage{bbm}
\usepackage{epsfig}
\usepackage{amssymb,amsmath,amscd, mdwmath}
\usepackage{graphicx,cite,url}
\usepackage{algorithmic}
\usepackage{float}
\usepackage{amsfonts}
\usepackage{syntonly}
\usepackage{multirow}
\usepackage{graphicx}
\usepackage{subfig}
\usepackage{tabularx}
\usepackage{setspace}
\usepackage{stfloats}
\usepackage{amsthm}

\usepackage[labelfont=bf,labelsep=period,justification=raggedright]{caption}

\makeatletter
\renewcommand{\@biblabel}[1]{\quad#1.}
\makeatother

\date{}

\begin{document}

\begin{flushleft}
{\Large
\textbf{Predicting publication productivity for researchers: A latent variable model
 }
}
\\
Zheng Xie$^{1,2, \sharp }$
\\
\bf{1}   College of Liberal Arts and Sciences,   National University of Defense Technology, Changsha,   China. \\
\bf{2} Department of
Mathematics, University of California, Los Angeles,   USA
\\  $^\sharp$ xiezheng81@nudt.edu.cn
 \end{flushleft}
\section*{Abstract}
This study provided a model for the publication dynamics of researchers, which is based on the    relationship between the publication productivity of researchers and two covariates: time and historical publication quantity. The relationship allows to estimate the latent variable the publication creativity of researchers.
The variable is  applied to the prediction of  publication productivity for researchers. The
  statistical significance of the relationship
is validated by the high quality dblp dataset.
The
effectiveness of the model is testified on the dataset  by
   the fine   fittings on
    the quantitative distribution of researchers'  publications, the evolutionary trend of their publication productivity, and the occurrence   of  publication events.
  Due to its nature of regression,  the model has the potential to be extended for assessing the confidence level of prediction results, and thus has applicability to empirical research.





%


\noindent {\bf Keywords:}  Scientific publication,  Productivity prediction, Data modelling.

\section*{Introduction}

Predicting the scientific success of  researchers  is a   topic in the scientific fields such as informetrics, scientometrics, and bibliometrics\cite{Sinatra2016}.
It has three main  aspects: the $h$-index\cite{Hirsch2005},     citation-based indexes, and    publication productivity.
Much attention on this topic has concentrated  on   the $h$-index:  the maximum value of $h$ such that a researcher   has produced $h$ publications that have each been cited at least $h$ times.
The  popularity of the $h$-index is attributable  to its simplicity and its addressing both the productivity and the citation impact of  publications\cite{Schubert2007}.
  Acuna et al presented a prediction formula to  show that the current $h$-index is the most significant   predictor, compared
with the number of current papers, the year  since publishing first paper, etc\cite{Acuna2012}.
 Mccarty  et al   showed  that the number of
 coauthors and their  $h$-index also are significant  predictors\cite{Mccarty2013}.
The formula provided by
  Dong
et al   utilized    more  features, such as  the average  citations of an author's papers, and the number of coauthors\cite{Dong2016}.


 Predicting highly cited publications by short-term citation data also attracts considerable attention. Mazloumian applied a multi-level regression model\cite{Mazloumian2012}; Wang et al derived a mechanistic model\cite{Wang2013}; Newman defined $z$-scores\cite{Newman2014}; Gao et al utilized a Gaussian mixture model\cite{Cao2016}; Pobiedina applied link prediction\cite{Pobiedina2016}; Abrishami et al utilized deep learning\cite{Abrishami2019}. More information of publications (such as journal,  authors,  and content)  has been utilized to improve prediction precision:
    Bornmann et al considered
  publications' length\cite{Bornmann2014};
    Sarigol et al   used specific  characteristics of coauthorship networks\cite{Sarigol2014};
 Yu et al    synthesized   the  features of publications, authors, and   journals\cite{YuYu2014};
 Klimek et al utilized
the centrality measures of
term-document networks\cite{Klimek2016};
 Kosteas   considered the rankings of
 journals\cite{Kosteas2018}, Bai et al  used the aging  of   publications' impact\cite{Bai2019};
Stern and Abramo  utilized the impact factors of journals\cite{Stern2014,Abramo2019}.

Fewer attention  has concentrated  on
   the prediction of publication productivity,
compared with that on $h$-index and  citation-based indexes.  Empirical studies found the cumulative advantage in producing publications and the aging of researchers' creativity\cite{Newman2001,TomassiniM2007}. Laurance et al   found that  Pre-PhD publication success   strongly  correlates to long-term success\cite{Laurance2013}.  Lehman concluded that productivity usually begins in a researcher's 20s, rises sharply to a peak in
the late 30s or early 40s, and then declines slowly\cite{Lehman2017}. In the year 1984, Simonton provided a formula to model this process\cite{Simonton1984}, which, however, is at variance with the observations   on the vast numbers of current   researchers.

Aforementioned  prediction methods      of citation-based indexes and $h$-index
 all refer to the positive   correlation between the current indexes and their historical quantity.  The current $h$-index,
  the number of annual
citations,   and   the number of    five-year citations
  are  found to be   their positive predictors.
  The success of those methods
 can be thought to result from  the  cumulative advantage of receiving  citations found by Price in the year 1965\cite{Price1965}, which   has been extended  as a general theory for bibliometric and other cumulative advantage process\cite{Price1,Barabasi1999a,PercM2014}.
 From the perspective of   statistics,  the success is due to  the  predictable components  of these indicators    that can be extracted via autoregression.

The effect of cumulative advantage in producing  publications is weaker than that in receiving citations.
It is displayed  as    that
 the tail  of the quantitative distribution of the publications produced by a group of researchers   is much shorter than that of the citation distribution  of those researchers\cite{XieX2017}.
 Therefore, the critical factor of the   success in the prediction   of citation-based indexes and the $h$-index
does not exist   in the prediction   of publication productivity.
Then, is there still any  predictability in  publication patterns?









We provided a model to estimate
  the latent variable the creativity of researchers on   publications,   termed publication creativity,  through  an observed variable
  publication productivity. Given a group of researchers,
the quantitative distribution  of their  publications  can be  thought as      a mixture of   Poisson distributions\cite{Glanzel2,XieO2018, XieLL2018,   XieO2016}.
Samples following the same  Poisson distribution means that they would be   drawn from the same population. It means researchers can be partitioned into several populations, each of which  has certain homogeneity  in publication patterns.  And
 the publication creativity
of each subset of the partition
  can be regarded  as the expected value of the corresponding  Poisson distribution.
  Therefore, finding such a  partition will contribute   to revealing the mystery of publication patterns.


 We used  a
partition scheme, such that each subset consists of  the researchers with the same number of  historical publications produced before the current time.
When applying to  the  dblp dataset,   we found that
for the majority of researchers, the scheme satisfies our requirement.
Furthermore, we   found that researchers' publication  productivity   
    significantly correlates to time given a quantity of historical publications, and to researchers' historical publication quantity given a time.
It allows   us to estimate researchers'   publication creativity   at each time interval,
and then   to
   infer their productivity in the future.
 Three methods are provided to test the prediction results of our  model  in terms of  the evolutionary trend of productivity, the quantitative distribution of publications,  and the occurrence  of publication events.
 Compared  with the model in  the previous work\cite{Xie2019ppp}, the model removes
 the limitation on
the prediction for  high  productive researchers, however, its precision still needs to be improved. In addition,
the model  decreases the requirement of training dataset on  the quantity of productive researchers.




This paper is organized as follows.  The model and its motivation are described in Sections 2, 3. The empirical  data and experiments are described in Section  4. The results are discussed and conclusions drawn in Section 5.

\section*{Motivation}


\subsection*{Latent   publication creativity}

 Finding regularities in the
publication patterns of researchers is an elusive task. Several factors on individual level  have been used to explain the productivity levels among researchers, such as work habits,  psychological traits (intelligence, creativity, etc.), demographic characteristics (aging, gender, etc.), and environmental location (graduate school background, the prestige of institution, etc.)\cite{Stumpf1995}. Many of these investigations  focused on  the  psychological traits.
For example, Andrews reported that creativity may not result in productivity unless the researchers have strong motivations\cite{Andrews1976}.
Our study   considers the creativity  on producing publications,   termed    ``publication creativity".



A latent variable model  contains latent variables   that
cannot  be directly observed. Those variables  have connections to
  certain  variables that can be observed directly.
  The annual publication creativity  of a researcher  is  such an unobserved variable.
  The provided model gives a method to  measure it   as the expected value of
 an observed variable the
   annual publication productivity  of researchers.
 The  publication productivity of a researcher is   easily affected by random factors from his or her work environment, family, etc.
 Meanwhile, its expected value    could be relatively  stable, which would be more suitable for the prediction   of  publication quantity.

 \subsection*{Inhomogeneous  Poisson  process}
How to define the publication creativity of researchers, and how to calculate  it based on   empirical data?
The definition  given here is  based on the feature of the quantitative distributions  of  researchers' publications.
These   distributions   can be fitted by a mixture of Poisson distributions\cite{XieO2016}. Therefore, we could expect to partition researchers into specific  subsets, such that the   distribution of each subset    is a Poisson distribution.
In this study, the   publication creativity of any researcher in the subset   is defined to be  the expected value of the Poisson distribution.  The publication productivity  of the researcher is treated  as
a sample from the distribution; thus has certain randomness.



How to find those subsets?
 Previous studies show that
the   distributions
are featured by a trichotomy, comprising a generalized Poisson head, a power-law middle part, and an exponential cutoff\cite{Xie2019Scientometrics}.
 The trichotomy  can be derived from a range of ``coin flipping" behaviors, where the probability of observing ``head" is dependent on observed events~\cite{Consul}. The event of producing  a publication can be regarded as an analogy of observing ``head", where the probability of publishing is also affected by previous events. Researchers would easily produce  their later publications, compared with their first one. This is a cumulative advantage, research experiences accumulating in the process of producing publications. It displays as the transition from the generated Poisson head to the power-law part. Aging of researchers' creativity is against cumulative advantage,   displaying as the transition from the power-law part to the exponential cutoff.

When considering a short time interval, the effects of cumulative advantage and  aging would be not significant. However, the diversity of researchers in publication history cannot be eliminated only by shrinking the observation window in the time dimension. Therefore,
we
partitioned
  researchers into  $I$ subsets, where $I$ is a given integer. For    $i=1,2,...,I$,
   the $i$-th subset  contains the researchers with $i$ historical  publications before the current time.
   In practice, the historical publications are counted since   a given time $T_0$.

\subsection*{The Lotka's law}
Lotka  analyzed the publications of   physics journals during the nineteenth century, and found  the  law: the publication quantity  of a researcher approximately satisfies  that the number producing $n$ (where $n\in \mathrm{Z}^+$) publications is about $1/n^2$ of those producing
one\cite{Lotka1926}.  It is named  Lotka's law and stimulates
Price's  study  on  the patterns of long-term historical  publications.   Price provided   his  ``inverse square law" that   half of the publications come from the square root of all researchers\cite{Price1963}.

Lotka's law is now usually
defined in the generalized form  $p(x=h)\propto h^a$,
where $a<-1$,  $h\in \mathbb{Z}$,   $x$ is random variable,
and
$p(x=h)$ represents the probability  that a researcher  produced  $h$  publications.
Therefore, given a set of researchers, the probability  of a researcher with $s$ publications at time interval $[T_0,T_1]$ can be supposed to be proportional to $s^a$.
Assume that the researcher's publication productivity at  a following time interval $(T_1,T_2]$    is $s^b$, where $b>0$.
 It gives rise to  $p(x=s^b)\propto s^a$ at time interval $(T_1,T_2]$. Letting $h=s^b$ obtains  $p(x=h)\propto h^{a/b}$.
 Therefore,   with the assumption,
 the Lotka's law holds at the following time interval. It  gives  certain reasonability to the assumption, which is  needed by the log-log regression of the model.

\section*{The    model}

\subsection*{Model    terms  }
Consider  the researchers
 who produced publications  at  two time intervals $[T_0,T_1]$ and $[T_1,T_2]$.
Our aim is to estimate researchers' publication  creativity    at   $[T_1,T_2]$.
Partition  $[T_1,T_2]$ into $J$ intervals with cutpoints $T_1 = t_0 < t_1 <
\cdots < t_J = T_2$.
   The half-closed interval  $(t_{j-1}, t_j]$ is referred to as  the $j$-th time interval, where $j=1,2,...,J$.
  Partition   the researchers with no more than $I$ publications at $[T_0,t_j]$ into  $I$ subsets
  according to  their historical publication quantity at $[T_0,t_{j-1}]$. That is,
  the $i$-th subset consists of the researchers with $i$   publications at $[T_0,t_{j-1}]$.

  Let the    $i$-th subset  at the $j$-th time interval   be
the subset of the researchers with $i$   publications at $[T_0,t_{j-1}]$.
Define  the   publication productivity of the $i$-th subset
 in the   dataset
 at  $(t_{j-1},t_{j}]  $ to be the average publication quantity of its researchers at that time interval, and denote it by $\eta_{ij}$.
It  can be calculated   as follows:
 \begin{equation}\eta_{ij}= \frac{m_{ij}}{ n_{ij}}, \label{eq1}
\end{equation}
  where  $n_{ij}$ is the number of  researchers with $i$ publications at   $[T_0,t_{j-1}]$, and $m_{ij}$ is the
   number of publications produced by those  researchers  at $(t_{j-1},t_j]$.
    Define
  the publication creativity of    the $i$-th subset  at the $j$-th time interval to be the expected value of  $\eta_{ij}$, and denote it by $\lambda_{ij}$.

\subsection*{Regression formulae}



The provide model is  the combination of a piecewise
Poisson regression and a log-log regression.
Firstly, treating  the index $i$ of $\lambda_{ij}$   as a dummy index,   we assumed  $\lambda_{i1}>0$   and
\begin{equation}\lambda_{ij}=\lambda_{i1}   \mathrm{e}^{{ \beta_i} ({t}_j-t_1)  }, \label{eq2}
\end{equation}
   where $\beta_i $  tunes the  effect of ${t}_j$. The proportionate $\mathrm{e}^{ {  \beta}_i {  t}_j  }$  changes the creativity.
   Taking logs in   Eq.~(\ref{eq2})
 obtains
\begin{equation}\log \lambda_{ij} = \alpha_i+\beta_i ( t_j-t_1),\label{eq3}
\end{equation} where $\alpha_i=\log \lambda_{i1}$.
 For each   subset $i$,
  Eq.~(\ref{eq3}) is a one-variable  Poisson   model (see its definition in Appendix A).
 Because
 for the majority of researchers in any subset  $i$,
 their   number of publications produced  at a short time interval (e.~g. a year)
  follows a Poisson distribution\cite{Xie2019ppp}.




Secondly,
treating  the index $j$ of $\lambda_{ij}$   as a dummy index,   we assumed  $\lambda_{1j}>0$ 
 and \begin{equation}\lambda_{ij}=\lambda_{1j}  i^{\nu_j} , \label{eq4}
\end{equation}
   where  
$\nu_j $ tunes the effect  of $i$.
  Taking logs in     Eq.~(\ref{eq4})  obtains
\begin{equation}\log  \lambda_{ij} = \mu_j+\nu_j\log i,\label{eq5}
\end{equation} where $\mu_j=\log \lambda_{1j}$, and $\lambda_{1j}>0$ for empirical data.
Therefore, for each  time interval,
Eq.~(\ref{eq5})  is a one-variable  log-log   model (see its definition in Appendix A).

Except for the Poisson feature of empirical distributions of researchers' publications and the Lotka's law,
the reasonability of the assumptions of the regression formulae in Eqs.~(\ref{eq3}) and (\ref{eq5})
are also  based on the  significant  relationship
  between covariates and response variables  found  in  empirical data. Note that
the regression  formulae can be generalized to deal several  characteristics varying with $i$ and $j$. This study only considered the simplest case: one
characteristic for each index.


\subsection*{Training process}
\begin{figure*}[h]
\centering
\includegraphics[height=2.2  in,width=5.9     in,angle=0]{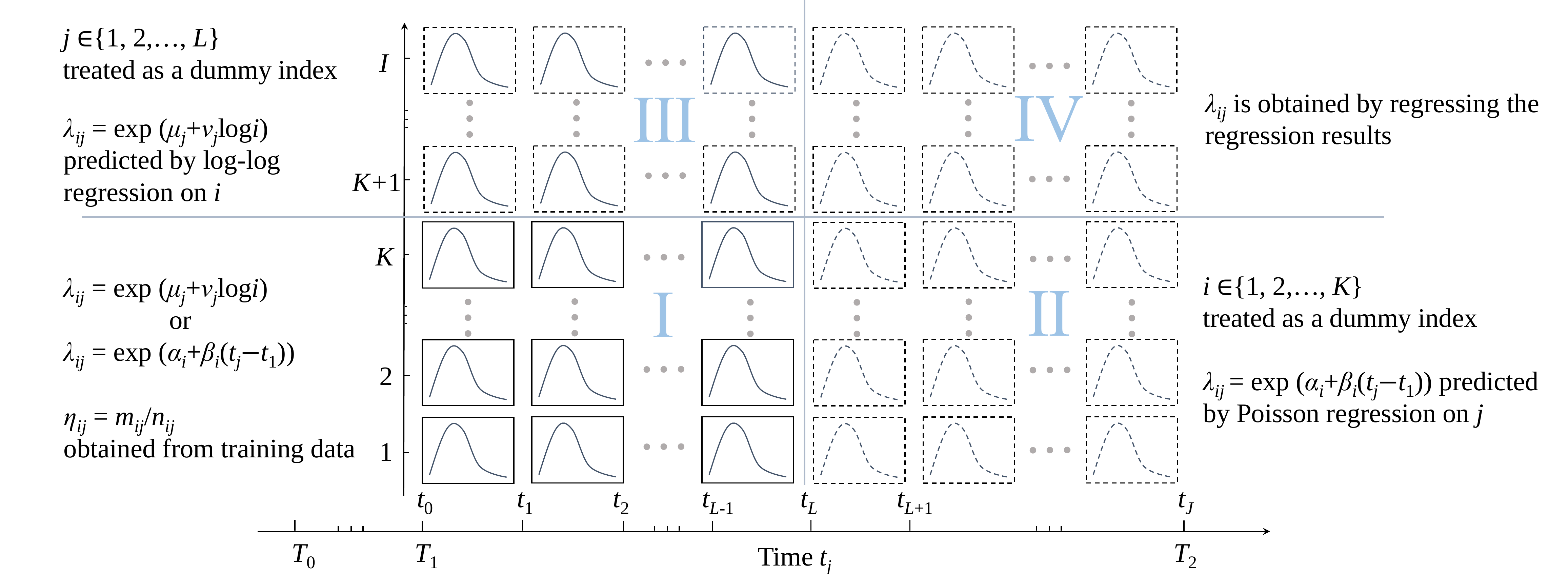}
\caption{   {\bf   An illustration of the latent model.}
 The publication productivity $\eta_{ij}$ in Zone I is calculated based on the training dataset.
  The matrix of publication creativity $(\lambda_{ij})_{I\times J}$ is divided into four zones,
which is calculated by the approach   shown in Section 3.
}
 \label{fig1}
\end{figure*}

 Consider a training dataset
consisting of the researchers producing  publications at the  time interval $[T_0, t_{L-1}]$  and their publications at the time interval $[T_0, t_{L}]$, where $L$ is an integer larger than $1$. At each time $t_l$ ($0<l<L$),
we considered the researchers of the training dataset, whose publication quantity at $[T_0, t_{l-1}]$ is no more than a given integer $K$.
Partition these researchers into $K$ subsets according to their  publication quantity at $[T_0, t_{l-1}]$.
 The training dataset is utilized to
calculate $m_{ij}$ and $n_{ij}$, namely the publication productivity  $\eta_{ij}$.
 The publication creativity can be expressed by  a  matrix $(\lambda_{ij})_{I\times J}$. To   show its computational process  clearly,
we divided the matrix    into four zones (Fig.~\ref{fig1}).

Firstly, we replaced the $\lambda_{ij}$ in Eq.~(\ref{eq3}) by $\eta_{ij}$, and  obtained
 \begin{equation}
  \log  \eta_{ij}  = \alpha_i+\beta_i (t_j-t_1).\label{eq6}
\end{equation}
Utilize the linear regression     to calculate $ \alpha_i$ and $\beta_i$   for   $i=1,...,K$.
Let $\lambda_{ij}=\mathrm{e}^{\alpha_i+\beta_i (t_j-t_1)}$, where  $j=1,...,J$.  It values the $\lambda_{ij}$
in Zones I and II.

Secondly, we replaced the $\lambda_{ij}$  in Eq.~(\ref{eq5}) by $\eta_{ij}$, and obtained
  \begin{equation}
 \log  {\eta_{ij}}   = \mu_j+\nu_j i.\label{eq7}
\end{equation}
Utilize the linear regression   to calculate  $ \mu_j$ and $\nu_j$  for  $j=1,...,L$.
 Let
   $\lambda_{ij}=\mathrm{e}^{\mu_j}i^{\nu_j}$, where $i=1,...,I$.   It values the $\lambda_{ij}$
 in Zones I and III.


Thirdly, we
substituted   the calculated  $\lambda_{ij}$ in Zone III into Eq.~(\ref{eq3}), and
  utilized  the linear regression      to calculate $ \alpha_i$ and $\beta_i$ for $i=K+1,...,I$. 
   Let $\lambda_{ij}=\mathrm{e}^{\alpha_i+\beta_i (t_j-t_1)}$,   where $j=L+1,...,J$. It values the  $\lambda_{ij}$ in Zone IV.
 We can also substitute    the calculated   $\lambda_{ij}$ in Zone II  into Eq.~(\ref{eq5}), and
  utilized  the linear regression     to calculate $ \mu_j$ and $\nu_j$ for $j=L+1,...,J$. 
   Let $\lambda_{ij}=   \mathrm{e}^{\mu_j}i^{\nu_j}$, where  $i=K+1,...,I$.  It also values the $\lambda_{ij}$ in Zone IV.
  
  Note that
the production creativity $\lambda_{ij}$  in Zones I and IV has two values.
 For the majority researchers of the training  dataset used in the next section, we  found that
$ \eta_{ij}$
 significantly correlates to $t_j$ given $i$ and to $i$ given $t_j$.
Therefore,  either the  two values of  $\lambda_{ij}$ or their average can be used    due to the statistical significance of  regression.


\subsection*{Predicting  publication quantity}

 Algorithm~\ref{tab1} is provided to  predict researchers'  publication quantity    at the time interval  $[t_X,t_Y]$, where $t_0\leq t_X<t_Y\leq t_J$.
  Denote  the publication quantity   of researcher   $s$   at  $[T_{0},t_l]$ by $h_s(t_l)$.
The algorithm gives  $h_s (t_Y)$  the predicted  publication quantity of researcher $s$ at    $[T_0, t_Y]$.
  Due to its regression nature,  the algorithm cannot exactly predict the publication quantity  for  an individual, but it can be suitable  for a group of researchers.


 Note that the training dataset would contain not  enough productive researchers. It would cause   the parameter $K$  much smaller than the largest  publication quantity $I$ that can be
 predicted by our model.
In this case, the model will give   bad prediction results
to productive researchers.




\begin{algorithm}
\caption{Predicting researchers'   publication quantity.}
\label{tab1}
\begin{algorithmic}
\REQUIRE ~~\\ 
the $h_s(t_X)$ of any   researcher $s$ in a test dataset;
\\the   publication creativity  matrix $(\lambda_{ij})_{I\times J}$. 
\ENSURE ~~\\ 
the predicted productivity  $h_s (t_Y)$ of   researcher $s$.
\FOR{each researcher $s$}
\STATE{initialize $h=h_s(t_X)$;}
\FOR{$l$ from ${X+1}$ to $Y$}
\STATE{sample an integer  $r$ from    $\mathrm{Pois}(\lambda_{hl})$; }
\STATE{let  $h= h    + r$;}
\ENDFOR
\STATE{let $h_s (t_Y)=h$;}
\ENDFOR
\end{algorithmic}
 \end{algorithm}






\section*{Experiments}

\subsection*{Empirical  data}

 To guarantee the prediction precision of the provided  model, we   need  a large  training dataset, which should contain  enough productive researchers. The dataset provided
by
the dblp computer science bibliography     satisfies our requirement, which consists of the open bibliographic information on the major journals and proceedings of computer science (https://dblp.org). The dataset is of high quality, because it has been corrected by several methods of name disambiguation and checked manually.



To show the improvement of the provided model,
 we used the some training   and test  datasets used  in Reference\cite{Xie2019ppp}  (see Table~\ref{tab2}).
  Sets 1 and 2 are used to extract the historical publication quantity for the test researchers in  Sets 3 and 4.
Sets 5 are 6 are used as training datasets.
Sets 7 and 8 are used to testify the prediction results for the researchers in  Sets 3 and 4.
   These datasets   consist of 220,344 publications  in 1,586 journals and proceedings, which are produced by  328,690 researchers at the years from 1951 to 2018.
   Due to the size and   time span
 of the analyzed  datasets,  
our model is at least   suitable for the    community of computer science.

\begin{table}[!ht] \centering \caption{{\bf Certain subsets  of the dblp   dataset.} }
\begin{tabular}{l ccccccccccc} \hline
Dataset&  $a$   & $b$ &  $c$  & $d$ & $e$ & $f$    \\ \hline
Set 1 &1951--1994 &   180,45& 18,398& 319& 1.558 &1.528 \\
 Set  2   &1951--2000 &  38,149& 35,643& 542 &1.571& 1.681\\
   Set   3   &1994 &2,903  &  1,922& 146 &  1.137  &  1.718 \\
 Set  4  &2000 & 5,741   &   3,600& 257  & 1.184 & 1.888 \\
 Set  5 & 1994--2009 & 88,853&64,558& 940& 1.545 &    2.126\\
 Set  6 & 1995--2009 & 87,140&62,636& 931& 1.538&     2.139\\
  Set   7 &  1995--2018 &316,212& 201,946 &1,538  & 1.754& 2.746\\
 Set   8 &  2001--2018 &  301,741& 184,701 & 1,495 &1.733& 2.831 \\
\hline
 \end{tabular}
  \begin{flushleft} The index  $a$:    the time interval of data,  $b$: the number  of researchers, $c$: the number of publications,  $d$: the  number of journals,  $e$: the average number of researchers' publications, $f$: the average number of publications' authors.
\end{flushleft}
\label{tab2}
\end{table}


In this section,
   Set   6 is used as the training dataset.  Its parameters are $I=180$,  $J=23$,  $K=42$, $L=14$,  $T_0=1951$, $T_1=t_0=1995$,    $t_L=2009$,  and $t_{J}=T_2=2018$.   
The test dataset  here  consists of the researchers  in Set 4, their historical publication quantity  from Set 2,
 and their annual publication quantity  from Set 8.   Its  parameters are $t_X=2000$, and  $t_Y=2018$.


The  publication productivity matrix  $(m_{ij}/n_{ij})_{K\times L}$   is   calculated  based on   the training dataset.
For example,
 $n_{11}$  is the number of researchers with one publication at   $1[951,1995]$, and
 $m_{11}$ is publication quantity  of those researchers   at the year $1996$.
 The publication creativity matrix $(\lambda_{ij})_{I\times J}$  is calculated by the method described in   Section 3.

 We   predicted the publication quantity  only for  99.98\% test  researchers  who have no more than $I_1=60$ publications at the time interval $[1951,2000]$.
   Note that    their predicted    publication quantity   can be  more than  $K=42$.
    That is, the model here removes a limitation of   the model in Reference\cite{Xie2019ppp}:
        the   upper  limit of  predicting  output  controlled by the parameter $K$.

\subsection*{The reasonability of model assumptions}
\begin{figure*}[h]
\centering
\includegraphics[height=3.   in,width=4.6     in,angle=0]{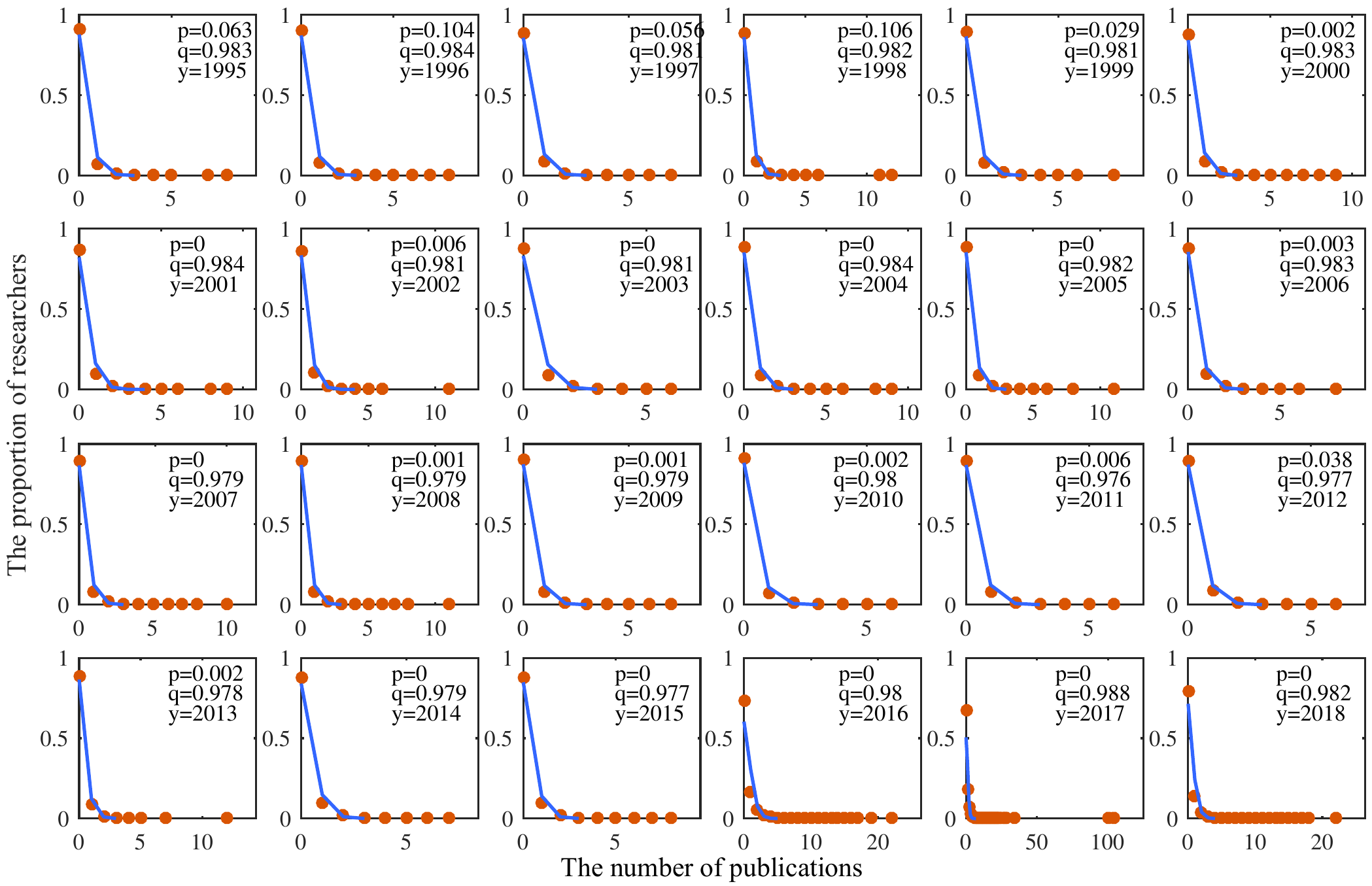}
\caption{    {\bf The quantitative distribution of researchers' publications.  }
Consider  the researchers  who produced  publications at  the year $y$
 and    no more than 10 publications at  $[1951,y]$. 
Index $q$ is their proportion  to the total    researchers who produced  publications at   $y$.
 When $p<0.05$, the KS test rejects  that
  the quantitative distribution of  the publications produced by the considered researchers   at the year $y+1$   (red circles)
is  a Poisson      (blue lines). }
 \label{fig2}
\end{figure*}

Firstly, we showed the Poisson nature of the      quantitative distributions of researchers' publications.
Consider the researchers  of the training dataset who have  publications at  the year $y$
 and    no more than 10 publications at $[1951,y]$, where $y=1995,...,2018$. Consider
the quantitative distributions of their publications produced at $y+1$.
 The   Kolmogorov-Smirnov (KS) test rejects to regard some of them as Poisson distributions  because of
  their     tail (Fig.~\ref{fig2}).
 



\begin{figure*}[h]
\centering
\includegraphics[height=3.   in,width=4.6     in,angle=0]{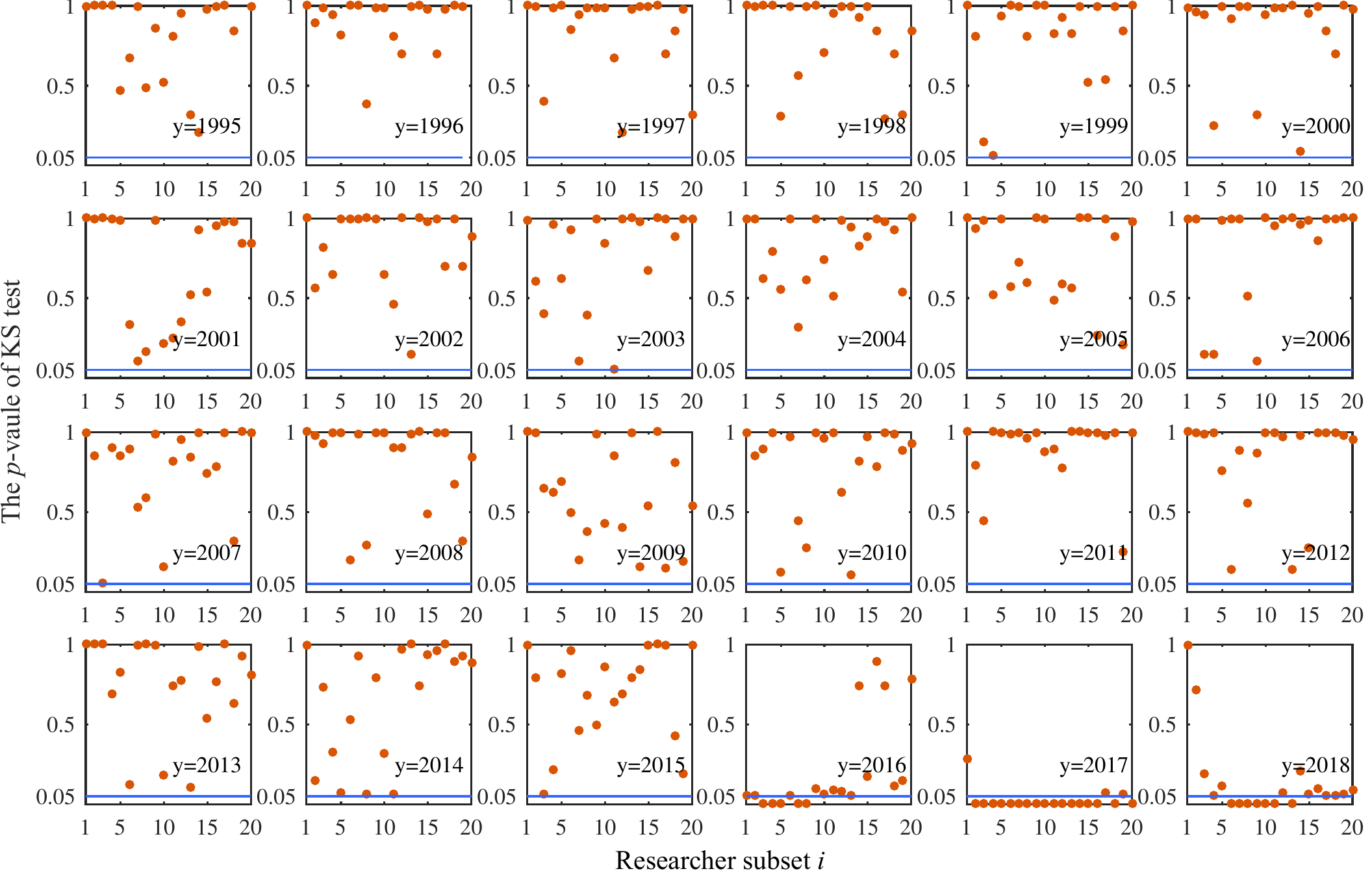}
\caption{    {\bf Eliminating the diversity in researchers'   historical publication quantity   induces    Poisson   distributions.  }
Consider the researchers who produced publications at $y$ and $i$ publications at $[1951,y]$.
When the $p$-value$>0.05$, the  KS test  cannot reject that  the quantitative distribution of 
the publications produced by these researchers at  $y+1$ is a Poisson. 
  }
 \label{fig3}
\end{figure*}

Consider   the researchers  of the training dataset who have  publications at  the year $y$
 and $i$ publications at $[1951,y]$. 
Fig.~\ref{fig3} shows that
   the quantitative distribution of their publications produced at $y+1$     is a Poisson, where $y=2001,...,2015$.
That is,
   diminishing the diversity in researchers'
historical publication quantity can reveal the Poisson nature of the  quantitative distributions of researchers' publications.
   
\begin{figure*}[h]
\centering
\includegraphics[height=0.9    in,width=2.3     in,angle=0]{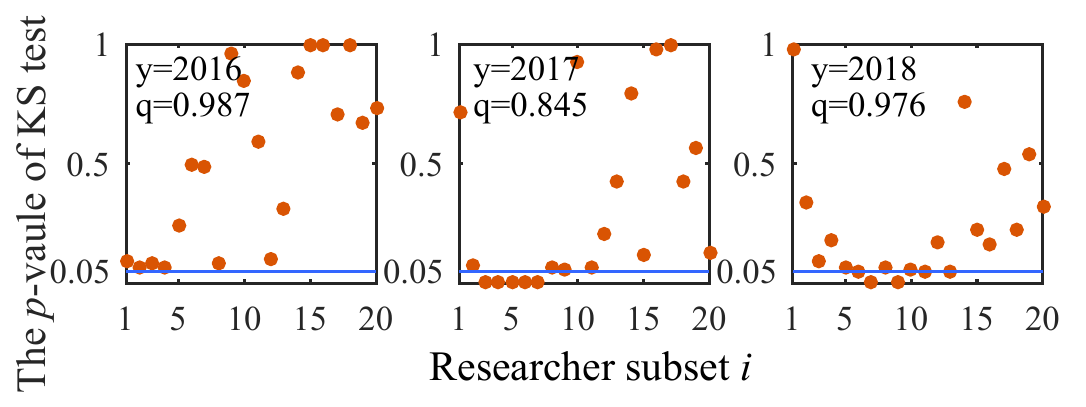}
\caption{    {\bf   Poisson feature holds for the majority of   researchers.  }
 The $p$-value is that of the KS test  in the caption of Fig~\ref{fig3}   on the   researchers  who produced   $i$ publications at $[1951,y]$,
   publications at $y$,
  and no more than $6$ publications at $y+1$.
Index $q$ is proportion of the researchers who    passed  the   test. }
 \label{fig12}
\end{figure*}

  There emerges a fraction of very productive  researchers at the year from 2016 to 2018. A few of them even produced  more than 100 publications a year, although their historical publication quantity is no more than 20.
  Consider the researchers who produced publications at $y$ and  no more than 6 publications at   $y+1$. Partition them into subsets according to their
  historical publication quantity at $[1951,y]$.
For some subsets of these  researchers, their publication distribution is  still   a Poisson  (Fig.~\ref{fig12}).
 It indicates that   the partition is suitable to calculate   publication creativity for
  the majority  of researchers  (see the proportion $q$    in Fig.~\ref{fig12});
 thus the principle of model still holds for those researchers.

\begin{figure*}[h]
\centering
\includegraphics[height=3.   in,width=4.6    in,angle=0]{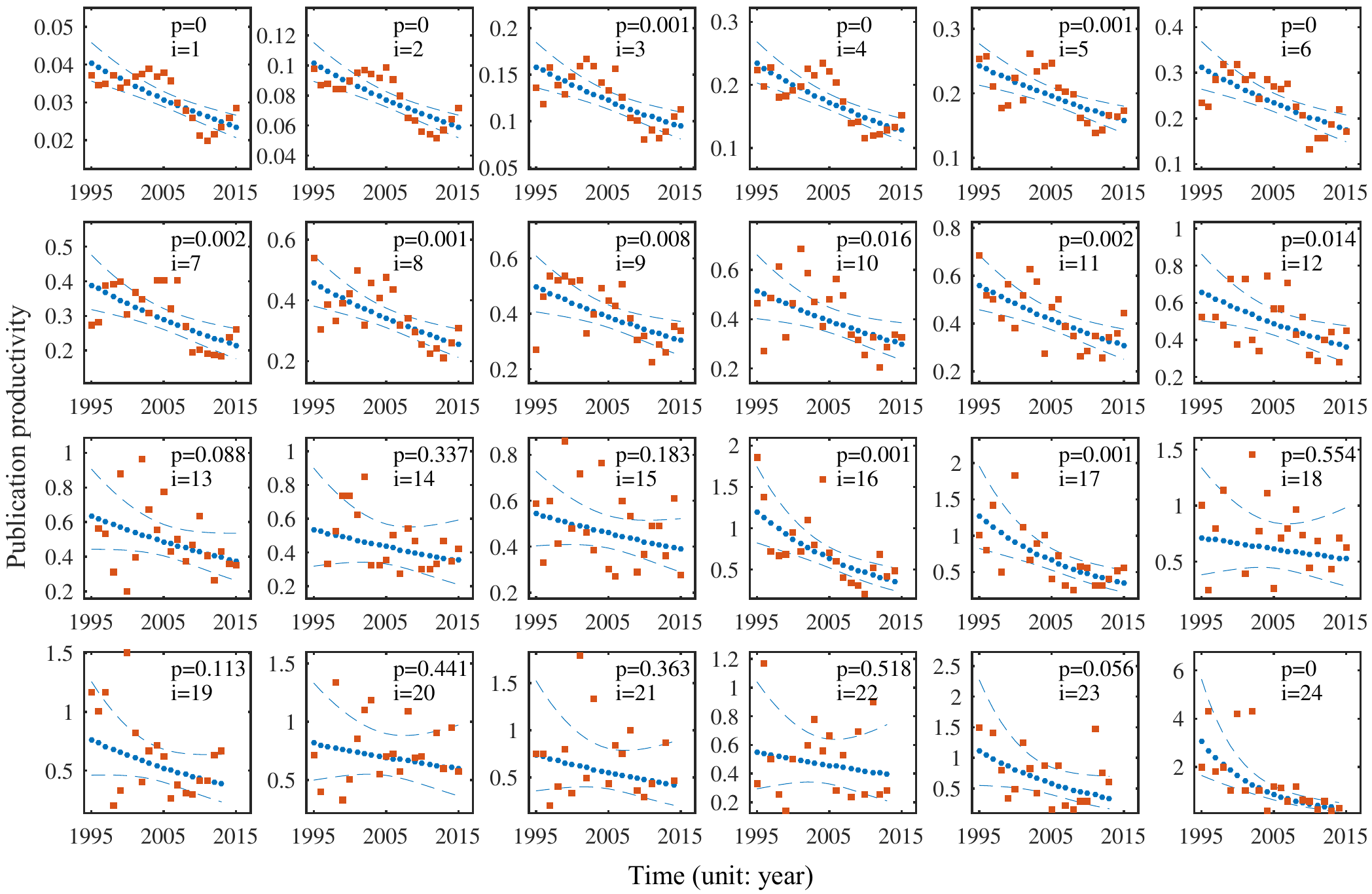}
\caption{   {\bf  The relationship between    publication productivity   and    time.}
 Consider the subset of the researchers    who have $i$  publications at  $[1951,y]$.     Panels show its  publication productivity at   $y+1$ (red squares),
   its predicted productivity  by the  Poisson model (blue dots), and  the confidence intervals of regression (dashed lines).
The relationship is significant,  when   $p<0.05$.  }
 \label{fig4}
\end{figure*}

\begin{figure*}[h]
\centering
\includegraphics[height=3.   in,width=4.6    in,angle=0]{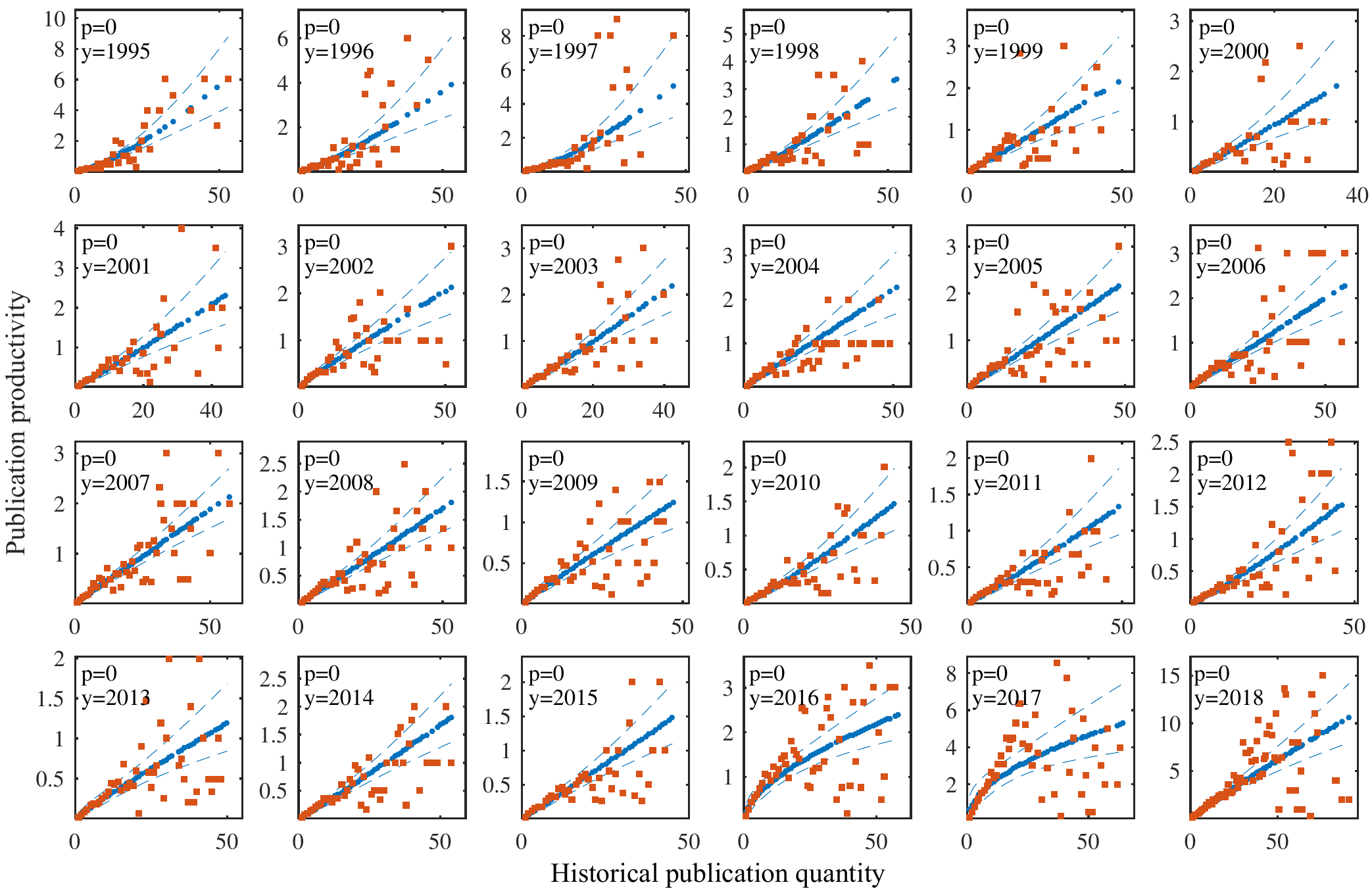}
\caption{   {\bf  The relationship between    publication productivity   and    historical publication quantity.}
  Consider the subset of the researchers    who have $i$  publications at  $[1951,y]$.   Panels show its  productivity  at the   year $y+1$ (red squares),
 its  predicted productivity by the  log-log model (blue dots), and   the confidence intervals of regression (dashed lines).
 When $p<0.05$,   the relationship    is significant.  }
 \label{fig5}
\end{figure*}

Secondly, we showed the significance  of regression    results on the training dataset. 
    The $\chi^2$ test indicates that  publication productivity 
  significantly correlates to    time
     given a historical publication quantity$\leq12$
     (see  the $p$-values in Fig.~\ref{fig4}). That is, the significance holds for $99.61\%$ researchers in the training dataset.
  The $\chi^2$ test  indicates that
  publication productivity  significantly correlates to historical publication quantity  given a time  (see  the $p$-values in Fig.~\ref{fig5}). 
The significance holds for all of the researchers in the training dataset.
 Those guarantee the effectiveness of calculating  $(\lambda_{ij})_{I\times J}$  by   regression.

\subsection*{Predicting  the evolutionary  trend  of publication quantity}

\begin{figure*}[h]
\centering
\includegraphics[height=2.2   in,width=4.6    in,angle=0]{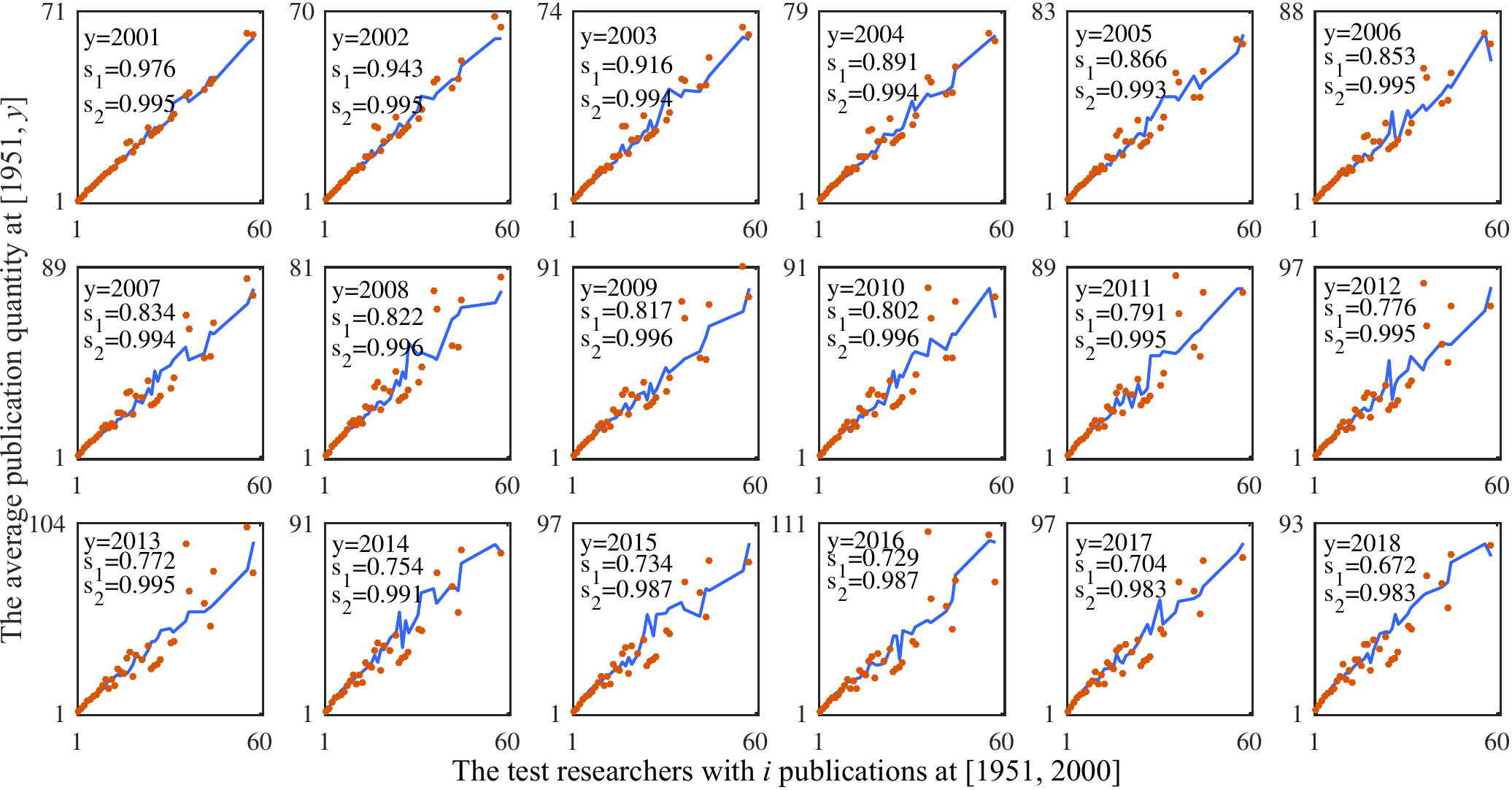}
\caption{   {\bf  Fittings on the evolutionary trend of   researchers' publication quantity.} Consider  the test researchers who have $i$ publications at $[1951,2000]$. 
 Panels show
  the average number of publications produced  by  these researchers at    $[1951,y]$ ($n(i,y)$,  red dots) and the predicted one    ($m(i,y)$, blue lines). 
 Index $s_1$ is the  Pearson  correlation coefficient
     calculated
 based on  the  list of researchers'   publication quantity and that of their predicted one.
   Index $s_2$ is that based on the sorted
    lists.}
 \label{fig6}
\end{figure*}

     Consider the test researchers   who produced $i$ publications at  $[1951,2000]$.
 Let $n(i,y)$ be the average  number of publications  produced  by these researchers at   $[1951,y]$, and $m(i,y)$ be the predicted one.
  Fig.~\ref{fig6} shows their   trend   about $i$  given
   $y$.
  The correlation between them is measured by the  Pearson  correlation coefficient\cite{Hollander} on individual level ($s_1$) and that on
 group level ($s_2$). Index $s_1$ decreases over time, whereas $s_2$ keeps high.
  It indicates that  the model is unapplicable to    the long-time prediction for an individual, but can be applicable for a group of researchers.


\subsection*{Predicting quantitative distributions of publications}
\begin{figure*}[h]
\centering
\includegraphics[height=2.2   in,width=4.6     in,angle=0]{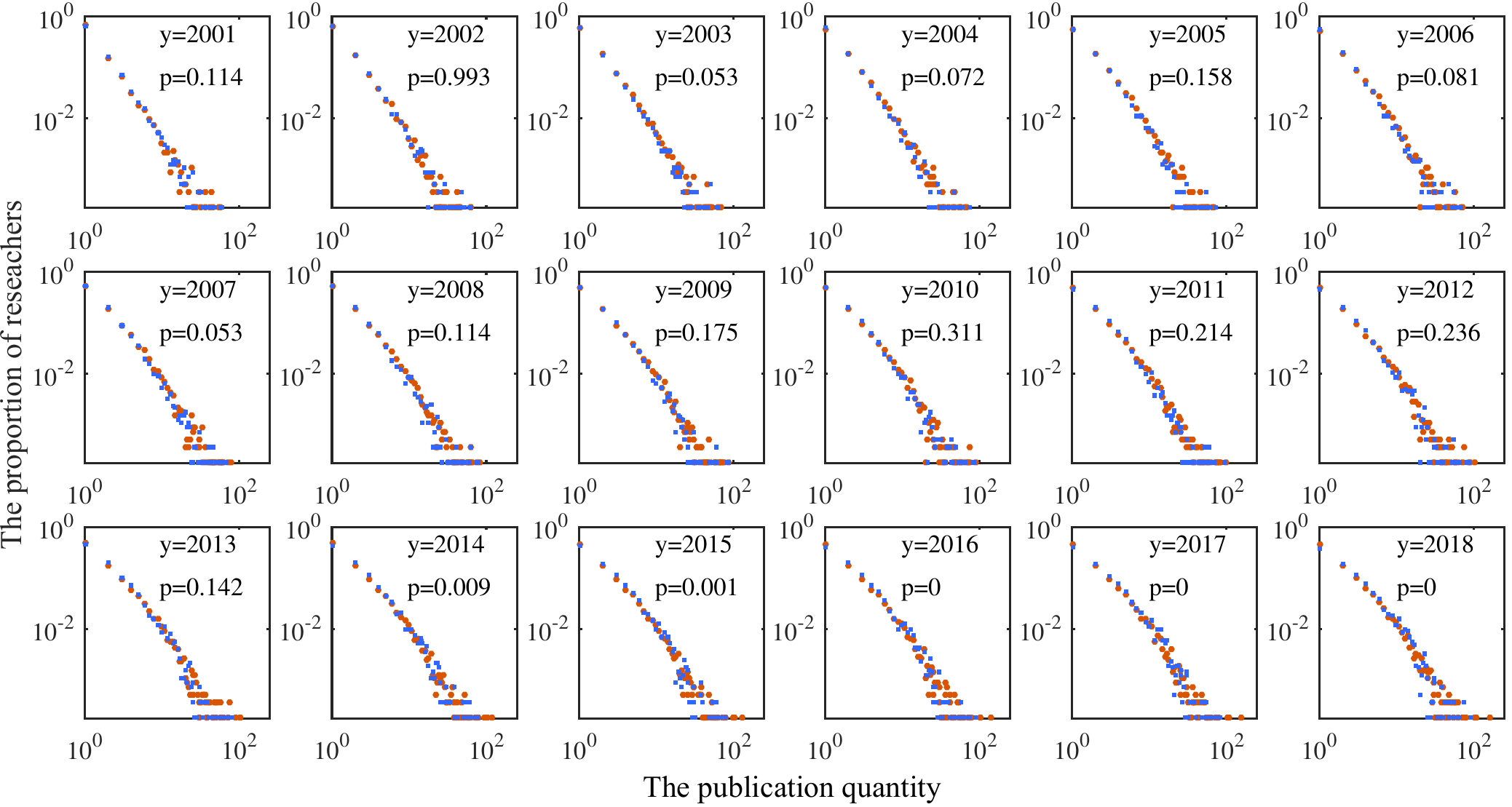}
\caption{   {\bf Fittings  on   the quantitative distribution of researchers' publications.}
Panels   show the quantitative distribution of
the  publications produced by the test  researchers
   at   $[1951 ,y]$ (red circles) and
  the  predicted one (blue squares).
When $p>0.05$, the KS test cannot reject the hypothesis that   the compared distributions are the same.
     }
 \label{fig7}
\end{figure*}

 We  compared the   distribution for the publications produced by   the test researchers at $[T_0,y]$   with the    predicted  one, where $y= 2001,..., 2018$. Fig.~\ref{fig7} shows that
a fat tail emerges in the evolution of
the  ground-truth distribution  and in that of the predicted one.
It shows that    our model can  capture the
fat-tail phenomenon.
Meanwhile, the predicted quantity  of the model in Reference\cite{Xie2019ppp} cannot be larger  than the  parameter $K$.
However, when the  time grows,  the KS test still rejects that the compared  distributions are the same (see the $p$-values in Fig.~\ref{fig7}), although there is a coincidence in their forepart. It indicates that the prediction precision   for productive researchers still needs to be improved.

\subsection*{Predicting publication events  }


The model gives  the publication creativity $\lambda_{ij}$
 to  the $i$-th subset  at the $j$-th time interval. It  can be used to calculate  $1-\mathrm{e}^{-\lambda_{ij}}$ the
 probability of  a test researcher (who has $i$ publications at $[T_0,t_{j-1}]$)   producing   publications  at $(t_{j-1},t_j]$.
The area under the receiver operating characteristic curve (AUC) is used  to measure   the   prediction precision on publication events.

Count the times when a   researcher who did (did not)    produce   publications at the next time interval
and the probability is larger (smaller)  than 0.5, and
denote the count by  $m_1$.
Count  the times when the probability is 0.5, and denote the count by $m_2$. Denote the number of tested researchers by $m$.
Index   AUC is calculated  as follows:
\begin{equation}\mathrm{AUC}= \frac{m_1+0.5m_2}{m}.\label{eq8}
\end{equation}

\begin{figure*}[h]
\centering
\includegraphics[height=2.2   in,width=4.6     in,angle=0]{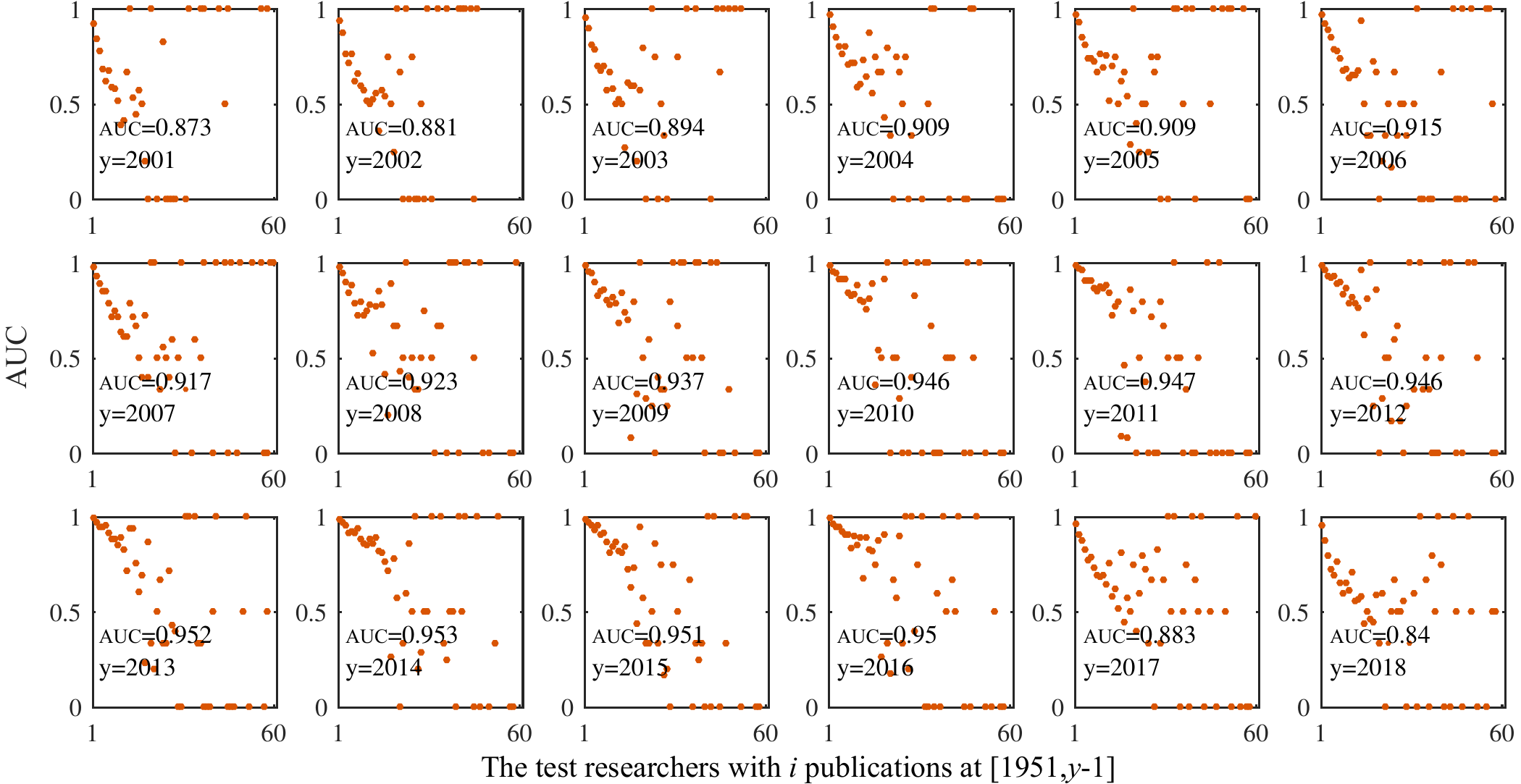}
 \caption{        {\bf  The precision of  predicting  publication events.} Consider   the test researchers   who produced $i$ publications at $[1951, y-1]$.
 Panels show
   the   AUC calculated by the formula in Eq.~(\ref{eq8}), which is the   precision
     of  predicting the publication   events
   at the   year $y$
    for  these researchers (red dots).  Index AUC is calculated based on   all of the test
researchers.         }
    \label{fig8}
\end{figure*}

Different from above two experiments focusing on the  prediction precision at a long time interval,
this experiment is designed  to measure the   precision at a short time interval, namely the next time interval.
Fig.~\ref{fig8}   shows that   index AUC is   high  on the researchers with a  small historical publication quantity $i$, which indicates the high precision  of predicting publication events   for low productive researchers.  It also shows there is no regularity can be revealed by the model for   productive researchers, which gives  the improving direction of the model. Due to the
vast number of low productive researchers,
the value of AUC is high   on  all of the test
researchers.


\section*{Discussion and conclusions}
We provided a model to estimate the publication creativity of researchers based on the
    relationship between
publication  productivity and two covariates, namely time  and
historical publication quantity.  The model offers convincing evidence that the publication patterns of the majority of  researchers are characterized by a piecewise Poisson process.
 Albeit simple, the   model could be specified
such that the  creativity can be formulated as a function  of time and the number of historical publications  with
 the coefficients estimated from data,
indicating the degree to which there is predictability.

The model is used to predict the  publication quantities    of the researchers in the dblp dataset.
Its predictability is testified  by the  fine fittings on the
evolutionary trend of researchers' productivity, the quantitative distribution of their publications,  and the probability of producing  publications.
   Compared with the model in Reference\cite{Xie2019ppp}, our model removes the  limitation  on predicting high productivity by using the log-log
  regression to estimate the publication creativity  of productive researchers, and thus decreases the requirement of training dataset on the quantity of productive researchers.

 Even where it does not provide an exact productivity prediction for     individuals, especially for productive  researchers,
 our model may still be of use in its ability to provide a satisfactory prediction for a group of researchers on average. Therefore, its prediction results
 offer some comfort:
for an individual,
 rejecting  a paper may feel indiscriminate and unfair, but for a group, these factors seem to average out.
   In addition,    due to its  advantage of providing results in an unbiased way, our model can be useful for funding agencies  to     evaluate
   the possibility of   completing the
quantitative index of publications in  applications.

Phenomena studied in human behaviors are
usually quite complex. Yet,
little is known about the mechanisms governing the evolution   of  researchers' publication productivity, whilst our model
  renders evolution  trajectories   relatively predictable on average.
Predicting the productivity of productive researchers individually  would not be done only by regression  as this study did for a group of researchers, due to the randomness of an individual's research. Analyzing massive data  to track scientific careers would   help to advance our understanding of how researchers' productivity evolves. Therefore, advanced algorithms are needed to synthetically analyze  the     features extracted
from researchers' historical publications, eduction background, and published journals. Especially, we should consider
the network features of their coauthorship (degree, betweenness, centrality, etc.), because previous  studies showed that research  collaboration  contributes to scientific productivity\cite{ Lee2005,DuctorL2015,QiMZengA2017}.








 \section*{Acknowledgments} The author thanks   Professor Jinying Su in the
National University of Defense Technology for her helpful comments
and feedback. This work is supported by the   National Natural   Science Foundation of China (Grant No. 61773020) and  National Education Science Foundation of China (Grant No. DIA180383).

\section*{Appendix A: the Poisson and log-log models}
The Poisson model with one covariate is   a   generalized linear model  of regression analysis\cite{Nelder-Wedderburn1972}. It is used to model count data and contingency tables, thus has potential  to  predict publication productivity.
It assumes that the response variable $y$ follows a Poisson distribution, and that the logarithm of its expected value can be expressed by a linear function of the covariate. Let
   $  {x} \in \mathbb {R}$ be a   covariate, and $\beta \in \mathbb {R} $ be  the effect thereof. The Poisson model takes the form
\begin{equation}{ \log( {\mathbb{E}} (y|  {x} ))=\alpha +\beta     {x}},\label{eq9}
\end{equation} where $\alpha\in \mathbb {R}$, and ${\mathbb{E}} (y|  {x} )$ is the conditional expected value of $y$ given $  {x}$.

The   log-log model with one covariate  assumes that the logarithm of its expected value can be expressed by a linear function of the logarithm of the covariate.  The   model takes the form
\begin{equation}{ \log( {\mathbb{E}} (y|  {x} ))=\alpha +\beta  \log{x}}.\label{eq10} \end{equation}

\section*{Appendix B: An other example}

\begin{figure*}[h]
\centering
\includegraphics[height=3.   in,width=4.6     in,angle=0]{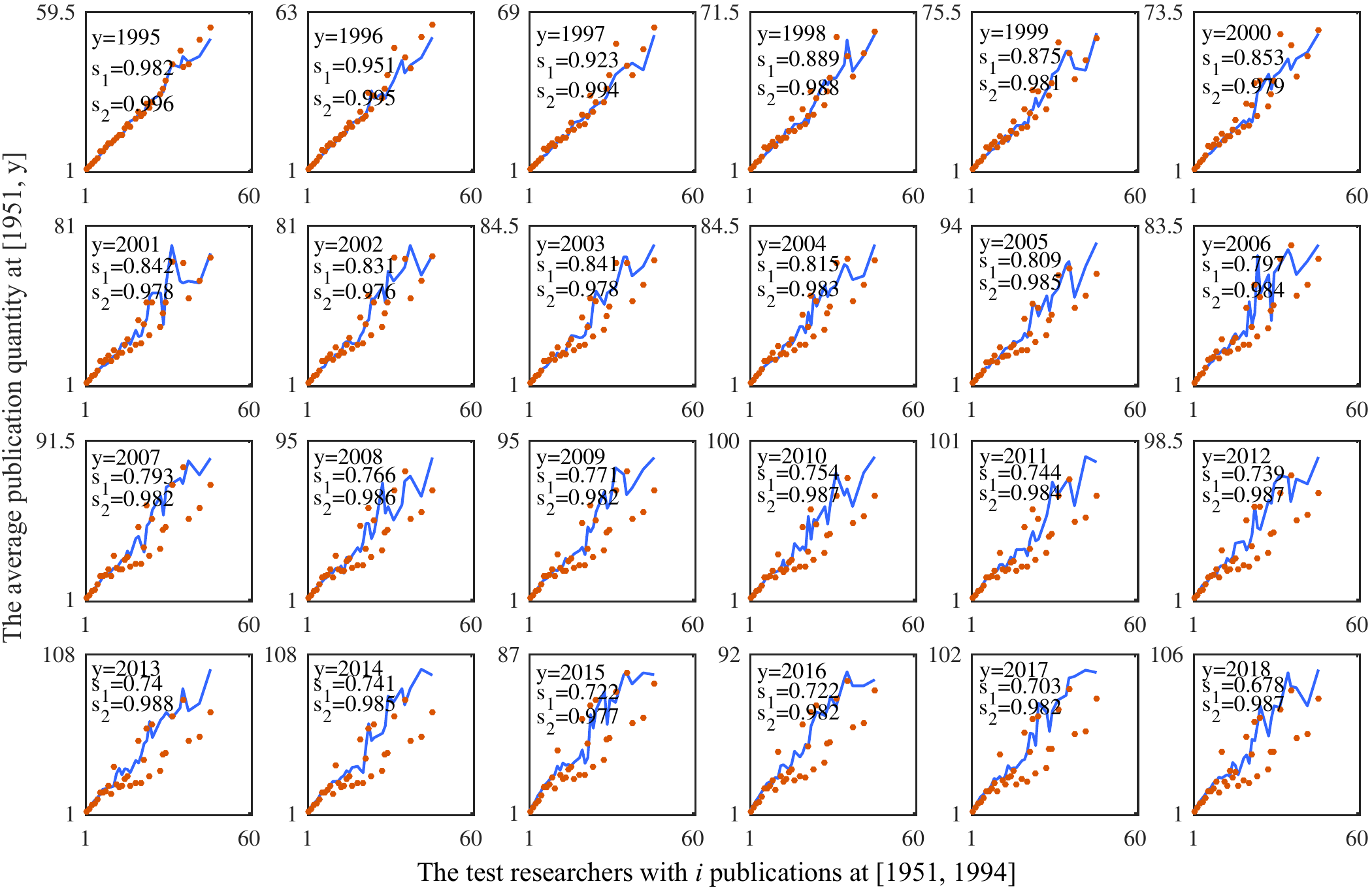}
   \caption{   {\bf  Fittings on the evolutionary trend of   researchers' publication quantity.} Consider  the test researchers who have $i$ publications at $[1951,1994]$.
 Panels show
  the average number of publications produced  by  these researchers at    $[1951,y]$ ($n(i,y)$,  red dots) and the predicted one    ($m(i,y)$, blue lines).
 Index $s_1$ is the  Pearson  correlation coefficient
     calculated
 based on  the  list of researchers'   publication quantity and that of their predicted one.
   Index $s_2$ is that based on the sorted
    lists.} 
    \label{fig9}
\end{figure*}

\begin{figure*}[h]
\centering
\includegraphics[height=3.   in,width=4.6    in,angle=0]{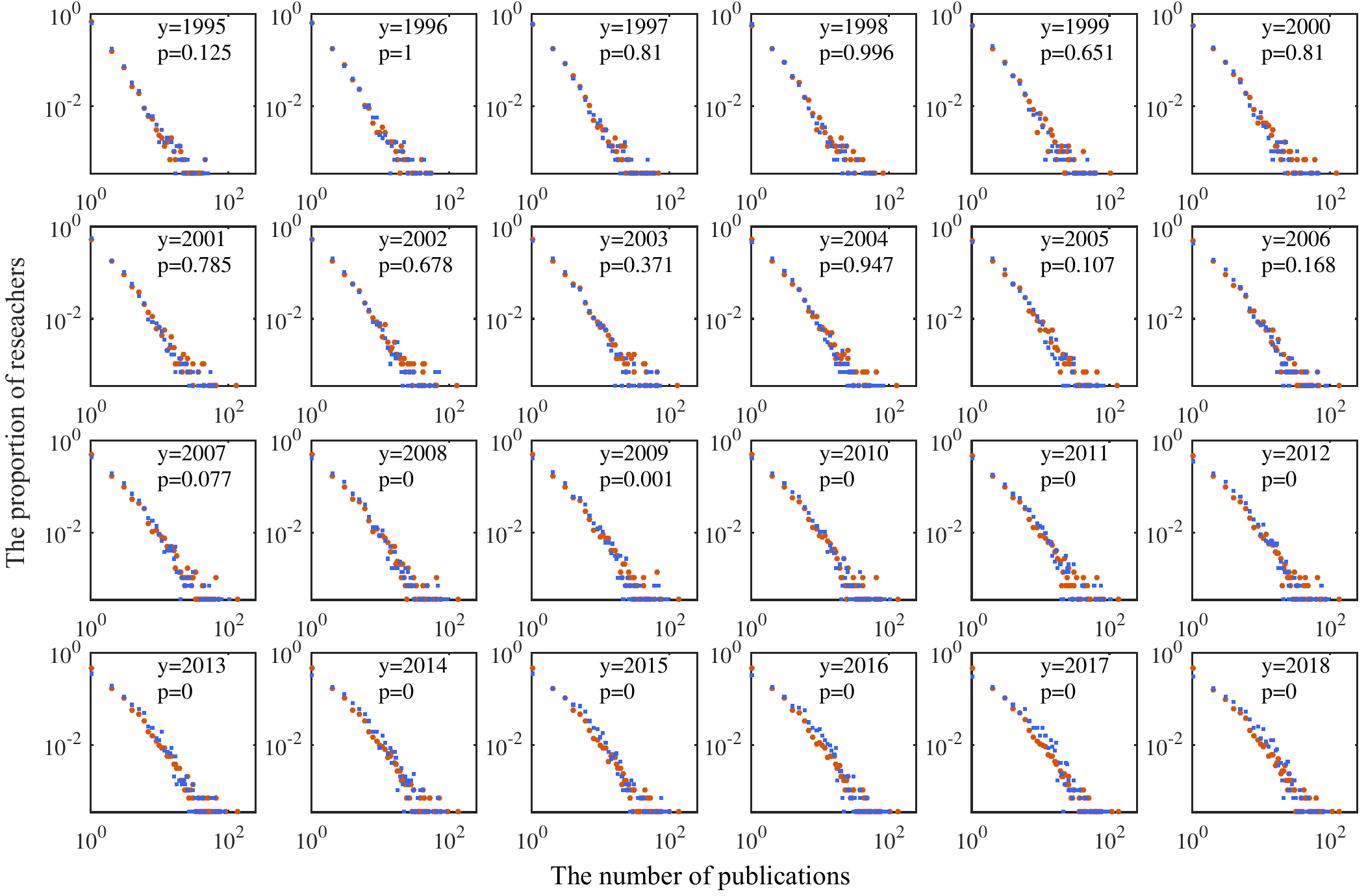}
  \caption{   {\bf Fittings  on   the quantitative distribution of researchers' publications.}
Panels   show the quantitative distribution of
the  publications produced by the test  researchers
   at   $[1951 ,y]$ (red circles) and
  the  predicted one (blue squares).
When $p>0.05$, the KS test cannot reject the hypothesis that   the compared distributions are the same.
     } 
    \label{fig10}
\end{figure*}

\begin{figure*}[h]
\centering
\includegraphics[height=3.   in,width=4.6   in,angle=0]{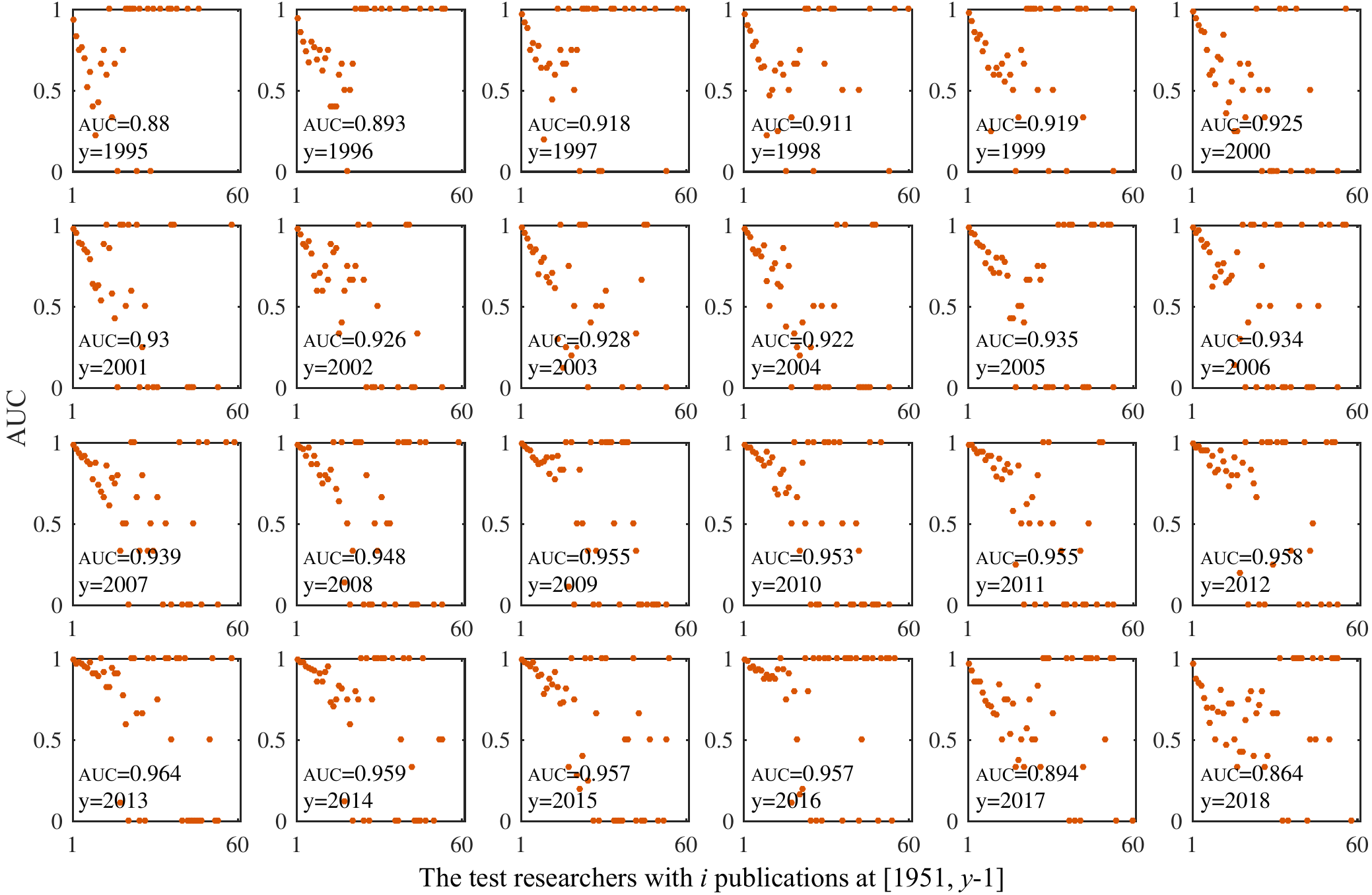}
 \caption{        {\bf  The precision of  predicting  publication events.} Consider   the test researchers   who produced $i$ publications at $[1951, y-1]$.
 Panels show
   the   AUC calculated by the formula in Eq.~(\ref{eq8}), which is the   precision
     of  predicting the publication   events
   at   $y$
    for  these researchers (red dots).  Index AUC is calculated based on   all of the test
researchers.         }
    \label{fig11}
\end{figure*}

   For    the experiment of this section,
    Set   5 is used as a training dataset. Its  parameters are
 $I=180$,  $J=24$,  $K=42$, $L=15$,  $T_0=1951$, $T_1=t_0=1994$,    $t_L=2009$,  and $t_{J}=T_2=2018$.  
 The test dataset here   consists of the researchers  from Set 3, their historical publication quantity from Set 1,
 and their annual  publication quantity in Set 7. Its  parameters are $t_X=1994$, and  $t_Y=2018$.
  We   only predicted the publications for    99.98\% test researchers  who have no more than $60$ publications at the time interval $[T_0,T_1]$.
Figs.~\ref{fig9}-\ref{fig11} show
the results of   the test methods  in Section 3.


\end{document}